%% file: main.tex
\documentclass[journal]{vgtc}
\onlineid{0}
\vgtccategory{Research}
\vgtcpapertype{please specify}

\graphicspath{{Figures/}{figs/}{figures/}{pictures/}{images/}{./}}
\usepackage{mathptmx}
\usepackage{amsmath} 

\title{Beyond the Broadcast: 
Enhancing VR Tennis Broadcasting through Embedded Visualizations and Camera Techniques}

\author{%
  Jun-Hsiang~Yao,
  Jielin~Feng,
  Xinfang~Tian,
  Kai~Xu,
  Gulshat~Amirkhanova, 
  and Siming~Chen
}

\authorfooter{
  \item
      Jun-Hsiang Yao, Jielin Feng, Xinfang Tian, and Siming Chen are with Ji Hua Laboratory and Fudan University. E-mail: rxyao24@m.fudan.edu.cn, simingchen@fudan.edu.cn. Siming Chen is the corresponding author.
      
  \item
       Kai Xu is with University of Nottingham. E-mail: kai.xu@nottingham.ac.uk. 
  \item
       Gulshat Amirkhanova is with Al-Farabi Kazakh National University. E-mail: gulshat.aa@gmail.com. 

}

\input{Section/0-Abstract}
\begin{document}
\input{Section/1-Introduction}
\input{Section/2-RelatedWork}
\input{Section/3-PreliminaryStudy}
\input{Section/4-Method}

\input{Section/5-Evaluation}
\input{Section/7-Limitation}

\input{Section/8-Conclusion}

\acknowledgments{%
    This work was supported by the Ji Hua Laboratory S\&T Program (X250881UG250) and Natural Science Foundation of China (NSFC No.62472099). The article was also supported by the project from the Ministry of Science and Higher Education of the Republic of Kazakhstan, No.BR24992975.
}

\bibliographystyle{abbrv-doi-hyperref}
\bibliography{main}

\end{document}

%% file: Section/0-Abstract.tex
\abstract{%
Virtual Reality (VR) broadcasting has emerged as a promising medium for providing immersive viewing experiences of major sports events such as tennis. 
However, current VR broadcast systems often lack an effective camera language and do not adequately incorporate dynamic, in-game visualizations, limiting viewer engagement and narrative clarity.
To address these limitations, we analyze 400 out-of-play segments from eight major tennis broadcasts to develop a tennis-specific design framework that effectively combines cinematic camera movements with embedded visualizations. We further refine our framework by examining 25 cinematic VR animations, comparing their camera techniques with traditional tennis broadcasts to identify key differences and inform adaptations for VR.
Based on data extracted from the broadcast videos, we reconstruct a simulated game that captures the players’ and ball’s motion and trajectories.
Leveraging this design framework and processing pipeline, we develope \emph{Beyond the Broadcast}, a VR tennis viewing system that integrates embedded visualizations with adaptive camera motions to construct a comprehensive and engaging narrative. Our system dynamically overlays tactical information and key match events onto the simulated environment, enhancing viewer comprehension and narrative engagement while ensuring perceptual immersion and viewing comfort.
A user study involving tennis viewers demonstrate that our approach outperforms traditional VR broadcasting methods in delivering an immersive, informative viewing experience.
}

\keywords{Tennis, VR Broadcasting, Embedded Visualization, Camera Motion, Immersive Experience}

\teaser{
  \centering
  \includegraphics[width=\linewidth, alt={A view of a city with buildings peeking out of the clouds.}]{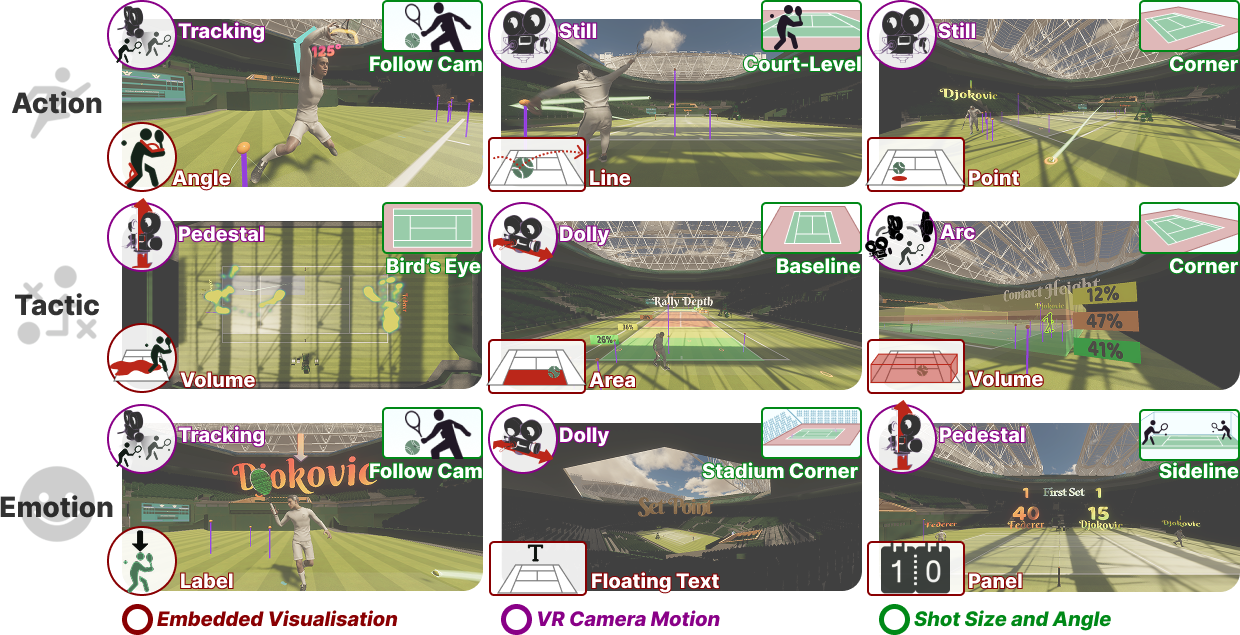}
  \caption{%
    Illustrative examples from our design framework demonstrate how individual shots combine \textit{Embedded Visualization, VR Camera Motion, and Shot size \& Angle} selections to support three \textit{Event} categories: Action (top row), Tactic (middle row), and Emotion (bottom row). Each panel presents a unique configuration—featuring elements such as dynamic tracking, baseline views, and text overlays—that integrates visual cues and camera techniques to effectively convey narrative elements in VR tennis broadcasting.
  }
  \label{fig:teaser}
}

%% file: Section/1-Introduction.tex
\firstsection{Introduction}\label{sec-1}
\maketitle


Sports broadcasting~\cite{buraimo2006demand} has undergone a dramatic transformation as emerging technologies reshape how audiences experience live events~\cite{solberg2011sports}. Traditional sports broadcasts, such as tennis~\cite{pingali1998real}, which has captivated audiences with significant popularity among individuals worldwide. The fast-paced and high-intensity nature of the game offers a unique blend of strategy~\cite{liu2024smartboard}, skill, and athleticism~\cite{pingali2000ball}. However, these conventional methods often fail to deliver the immersive, data-rich narratives that modern audiences demand~\cite{craig2015interactive}. 



Recent advances in Augmented Reality (AR)~\cite{soltani2020augmented} and Virtual Reality (VR)~\cite{kim2019impact} offer new opportunities to transform sports viewing~\cite{rematas2019watching, kim2019impact}. VR in particular, enables viewers to step virtually into the action with panoramic, stadium-scale perspectives such as NBA games or the Beijing Olympics~\cite{hayes2021immerse} that create a heightened sense of presence. 
This immersive capability~\cite{yao2025dtbia} brings audiences closer to the live event, creating a sense of presence that traditional broadcasts cannot match. 
In contrast, AR promises to enrich the viewing experience by overlaying rich game statistics and contextual information directly onto the real-world environment~\cite{du2024augmented,lo2023sports}, thereby deepening audience understanding without requiring them to divert their attention away from the field of play~\cite{bressa2021s}.

Despite their potential, both technologies face significant challenges. Current consumer AR devices are constrained by 3DoF tracking and real-time processing limitations, often resulting in static or poorly integrated overlays that do not adapt well to the dynamic nature of live sports~\cite{van2010survey,takrouri2022ar}. Meanwhile, existing VR broadcasts often rely on fixed camera perspectives, lacking dynamic motion and comprehensive narrative structure, thereby diminishing viewer engagement and coherence~\cite{zhu2023can}.
Embedded visualization techniques offer an effective solution by anchoring tactical and statistical information directly within its physical context, reducing cognitive load and minimizing attentional distractions~\cite{willett2016embedded,bressa2021s,lee2023design}. Previous studies successfully augmented traditional 2D broadcasts with embedded data overlays, enriching viewer understanding~\cite{zhu2023iball,chen2021augmenting,chen2022sporthesia}. However, integrating these approaches within dynamic, interactive VR scenarios remains challenging. Current VR broadcast frameworks generally feature static perspectives or isolated visualizations, limiting narrative coherence and the immersive potential offered by spatial integration and dynamic cinematography~\cite{zhu2023can}. Recent work indicates that integrating dynamic camera motions can significantly enhance narrative engagement, viewer immersion, and overall comprehension by guiding attention effectively~\cite{segel2010narrative,amini2016authoring}.
Consequently, it remains an open question on how best to integrate embedded visualizations with dynamic VR broadcasting techniques to fully satisfy in-game informational needs.

In this study, we address these challenges of translating traditional tennis broadcast language into an immersive VR narrative. Our investigation is guided by three key research questions. First, \textit{How can the cinematic language of conventional tennis broadcasts be adapted to enhance narrative engagement in VR?} To address this, we systematically annotated 400 out-of-play clips from eight major tennis tournaments to derive a design framework that captures how specific camera angles, shot sizes, and motion patterns contribute to the broadcast narrative. We further refined our design by analyzing 25 VR animation clips, comparing their camera usage with traditional tennis broadcasts to identify optimal adaptations for VR contexts. Second, \textit{How can dynamic camera movements and embedded visualizations be integrated to clearly communicate tactical and emotional match details in VR?} Leveraging our annotation insights, we developed \emph{Beyond the Broadcast}, an immersive VR tennis viewing system. Our system strategically integrates dynamic embedded visualizations with adaptive cinematic camera motions, constructing a coherent narrative to enhance viewer understanding of in-game tactics and emotional dynamics. Third, \textit{To what extent does this integrated approach enhance viewer comprehension and engagement without compromising immersion and comfort?} A controlled user study with 16 participants demonstrated that our integrated method significantly improved comprehension, narrative engagement, and perceptual immersion, all while maintaining viewer comfort and minimizing discomfort related to VR viewing.

\par Our contributions are summarized as follows:

\begin{itemize} 

    \item A systematic design framework derived from dual-annotations of 2D tennis broadcasts and cinematic VR animations, translating tennis-specific camera techniques into immersive VR narratives.
    
    \item We implement \emph{Beyond the Broadcast}, an end-to-end VR broadcasting system that transforms monocular tennis video inputs into immersive VR experiences, integrating embedded visualizations and adaptive camera techniques to facilitate intuitive narratives.
    
    \item Results findings from a controlled user study demonstrating the effectiveness of our integrated approach in improving viewer comprehension, narrative engagement, and perceptual immersion, while maintaining viewer comfort.
    
\end{itemize}

%% file: Section/2-RelatedWork.tex
\section{Related work}\label{sec-2}
This section reviews literature from three key domains: \textit{VR Sports Broadcasting}, \textit{Embedded Visualization in Sports}, and \textit{Cinematic Language for Virtual Reality}. 

\subsection{VR Sports Broadcast}
Sports broadcasting significantly expands the reach and appeal of live sports events to global audiences, with approximately half the world's population engaging through live broadcasts or highlights \cite{capasa2022virtual}. Recent technological advances, such as 360-degree cameras~\cite{inurrategi2008tv}, drones, and spidercams, have enriched viewer experiences by capturing diverse and dynamic angles~\cite{calagari2017sports}. Concurrently, advancements in VR and AR technologies have further transformed sports broadcasting. AR enhances on-site spectator engagement by overlaying real-time statistics, graphics, and interactive visual content onto physical environments, improving viewer understanding and interaction. VR, on the other hand, immerses remote viewers within simulated stadium settings, allowing exploration from various perspectives or athlete-specific viewpoints, significantly enhancing presence and viewer engagement~\cite{guo2024sports}.

Existing VR sports broadcasting methods predominantly rely either on panoramic videos captured by specialized 360-degree cameras or detailed 3D scene reconstructions derived from multi-view sensor arrays~\cite{wu2021construction}. Although these methods deliver high-quality immersive experiences, their dependence on specialized hardware and extensive production resources poses significant challenges in cost, scalability, and accessibility~\cite{yus2015multicamba}. To address these limitations, our research proposes reconstructing immersive 3D VR environments directly from monocular broadcast video footage commonly available online. This approach aims to provide an accessible, cost-effective solution for delivering engaging and immersive VR tennis viewing experiences.

\subsection{Embedded Visualization in Sports}
Embedded visualization integrates data representations directly within their spatial and temporal contexts to enhance viewer comprehension and engagement \cite{willett2016embedded}. In sports broadcasting, two predominant approaches have emerged: video-based overlays and immersive 3D scene integrations. Video-based overlays superimpose graphical elements onto traditional broadcast footage or AR displays, offering real-time contextual information \cite{lo2022stats, lo2023sports}. Systems such as iBall~\cite{zhu2023iball}, VisCommentator~\cite{chen2021augmenting}, and Sporthesia~\cite{chen2022sporthesia} enable analysts to create augmented sports content by overlaying motion trajectories, performance statistics, and tactical cues directly onto live video feeds. Early research by Pingali et al.~\cite{pingali2001visualization} demonstrated that such overlays could uncover nuanced insights into player performance and strategies—a concept refined in tennis by advanced systems such as LucentVision~\cite{dietrich2014baseball4d, stein2017bring}. Conversely, immersive VR environments facilitate fully reconstructed 3D scene integrations, wherein visualizations naturally align with their physical contexts~\cite{wagner2018virtualdesk}. Systems like Omnioculars~\cite{lin2022quest} and the immersive sports analytics framework by Guo et al.~\cite{guo2024sports} provide spatial insights beyond traditional overlays. Additionally, immersive analytics platforms like ShuttleSpace~\cite{ye2020shuttlespace} and TIVEE~\cite{chu2021tivee} effectively combine 2D data overlays with immersive 3D scenes, enriching real-time tactical analyses. The VIRD system~\cite{lin2023vird} exemplifies how dynamic, immersive visualization can enhance sports performance evaluation. Temporal factors also critically impact embedded visualization effectiveness. Static visualizations capture discrete, meaningful data points (e.g., player statistics or ball positions at specific match moments), whereas dynamic or "in-motion" visualizations continuously update with live match action, maintaining narrative coherence and contextual continuity~\cite{yao2022visualization, yao2024user}. Recent research on situated visualization~\cite{lee2023design, white2009sitelens} and immersive analytics~\cite{wen2022effects, weidinger2023immersive} confirms that embedding data directly within its physical context improves analytical clarity and enhances user engagement~\cite{srinivasan2024attention}. These techniques have been successfully applied across various sports disciplines—for example, real-time visual feedback in basketball free-throw training~\cite{lin2021towards}, mixed reality tactical guidance in basketball~\cite{cheng2024viscourt}, cycling performance analysis~\cite{kaplan2018towards}, and team sport tactical analytics~\cite{stein2017bring}.

In summary, while previous studies have separately advanced video overlay techniques~\cite{chen2021augmenting, chen2022sporthesia} and immersive 3D reconstruction methods~\cite{guo2024sports, lin2022quest}, a unified framework integrating both approaches specifically for VR sports broadcasting remains underexplored. Our work addresses this gap by combining both static and dynamic visualizations within an immersive VR tennis broadcast context. This integrated approach enhances narrative clarity, tactical insight, and viewer engagement, preserving the authentic spatial and temporal dimensions inherent to tennis~\cite{lo2023sports}.

\subsection{Cinematic Language for Virtual Reality}
Cinematic language has long relied on established conventions of framing, shot composition, and camera movement to guide audience attention and narrative clarity~\cite{lancaster2019basic,hodaie2017writing}. However, Cinematic Virtual Reality (CVR) introduces distinct challenges and opportunities, particularly due to the viewer's autonomy in controlling viewing angles~\cite{szita2018effects}. Matay and Bayar~\cite{matay2023cinematic} argue that CVR should be recognized as a unique narrative form, distinct from traditional cinema, requiring rethinking these established techniques to maintain narrative coherence and viewer comfort within immersive VR settings~\cite{mateer2017directing}. Recent research addresses core principles of effective camera usage in CVR. Passmore et al.~\cite{passmore2017360}, Rothe et al.~\cite{rothe2018impact,rothe2019spaceline}, and Keskinen et al.~\cite{keskinen2019effect} demonstrate that careful combinations of camera distances, shot scales, viewing angles, and interactions significantly enhance viewer comfort, immersion, and narrative effectiveness. Further studies highlight the importance of aligning camera motion with viewers' physical and perceptual expectations to mitigate disorientation~\cite{hebbel2017360,calagari2017sports,kohn2016you}. Additionally, Ross and Munt underscore the need to adapt screenwriting and narrative structures explicitly for spatialized VR storytelling, ensuring coherent viewer experiences~\cite{ross2018cinematic}. Zhang~\cite{zhang2020developing} propose specific methodologies and conceptual frameworks designed explicitly for CVR, advocating novel cinematic conventions tailored to the unique spatial characteristics of immersive 360-degree content. In sports broadcasting contexts, recent studies by Zhu et al.~\cite{zhu2023can} specifically evaluate the effectiveness of moving VR camera shots, while DreamStream~\cite{thoravi2022dreamstream} proposes interactive visualization strategies for immersive sports spectating.

Despite these advancements, there remains a notable gap in unified methodologies for integrating cinematic language and visualization, particularly in sports broadcasting contexts. Our research directly addresses this gap, proposing a comprehensive design framework that leverages these foundational theories to establish immersive and coherent cinematic narratives tailored explicitly for VR tennis broadcasting.

%% file: Section/3-PreliminaryStudy.tex
\begin{figure*}[htb]
     \includegraphics[width=\linewidth]{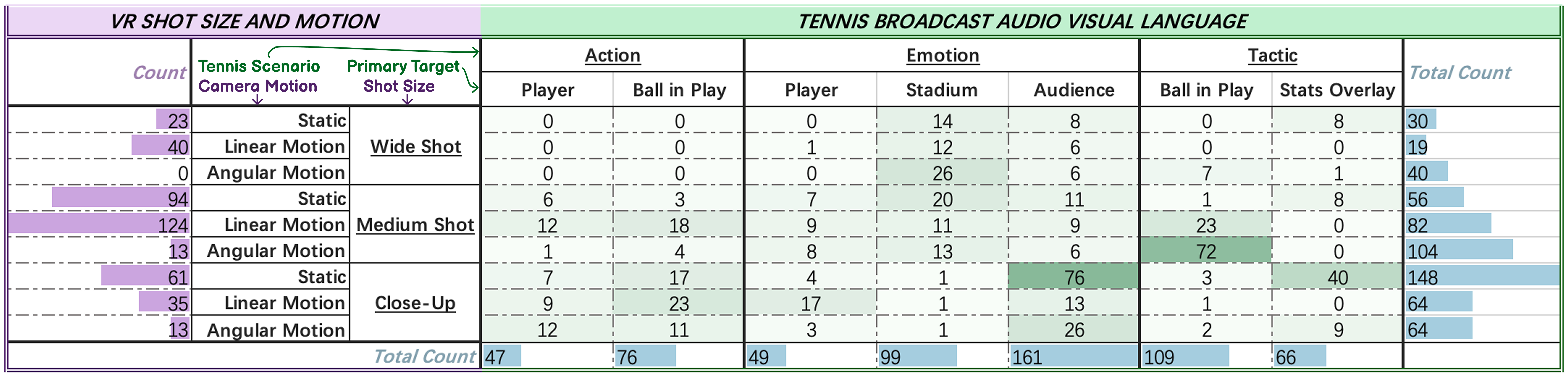}
    \caption{
    Comparative analysis of annotated camera usage in VR narrative animations \textit{(left, purple)} and traditional 2D tennis broadcasts \textit{(right, green)}. The right-side summarizes camera shot frequencies across narrative targets for three event categories, excluding repetitive angular pans focusing on player emotions. The left-side details corresponding VR annotations by shot size and camera motion type. Bar lengths represent normalized frequencies per side, with total counts provided. The comparison highlights differences: traditional broadcasts frequently utilize Close-Up shots and Angular motions to enhance emotional and tactical emphasis, whereas VR animations predominantly use Medium Shots with Static or Linear motions, optimizing narrative clarity and viewer comfort.
    }
     \label{fig:coding}
     \vspace{-0.5cm}
\end{figure*}

\section{Design Exploration}\label{sec-3}
This section introduces our dual-annotation approach, focusing first on conventional 2D tennis broadcasts and subsequently on cinematic 3D VR animations. 
Given the current lack of established cinematographic guidelines for VR tennis broadcasting, we aim to bridge this gap by comparing standard broadcast conventions with camera techniques specifically adapted for immersive VR contexts. 
We begin by examining traditional cinematographic methods used in tennis broadcasts, then analyze how these methods can be effectively adapted into VR environments. 
Finally, we synthesize insights from both analyses into a unified, coherent design framework.



\subsection{Tennis Specific Audio-Visual Language}\label{sec-3.1}
Audio-visual language~\cite{peng2017analysis} serves as the foundational expressive medium for all forms of video content. 
In sports broadcasting, and tennis in particular, mastering this language is critical to effectively convey the game's context, narrative dynamics, and emotional excitement to viewers~\cite{zhu2023can}. 
This language encompasses a set of cinematic techniques and conventions that are essential for conveying the context, narrative, and excitement of the game to the audience. 
In this subsection, we detail our systematic annotation process and present quantitative findings derived from traditional tennis broadcasts.

\textbf{Data Collection.} 
To examine the cinematic techniques employed in tennis broadcasts, we focused our analysis on the intervals between points, rather than the fixed baseline camera perspective typically used during active play. 
We selected eight high-profile Grand Slam matches (two each from the Australian Open, French Open, Wimbledon, and US Open) based on viewership, video clarity, and professional standards.
From each match, we extracted 50 consecutive out-of-play clips—totaling 400 segments—and segmented them into individual shots by detecting changes in framing or motion.
After removing 398 instances of the frequently occurring “Scoring Player / Non-Scoring Player + Close-up + Pan” combination—primarily used to depict players' emotional reactions, yet providing minimal variation—we obtained a refined dataset comprising 607 unique camera shot combinations for detailed analysis.
To systematically characterize each camera shot, we developed an annotation framework consisting of four key dimensions: \textit{Target}, \textit{Viewpoints}, \textit{Movements}, and \textit{Context}.
The \textit{Target} dimension identifies the primary focal elements of each shot, including scoring and non-scoring players, the ball in play, net and court areas, player boxes, statistical overlays, audiences, and stadium environments. Additional tags, such as "winner" or "forced error," provide supplementary context related to specific game events.
The \textit{Viewpoints} dimension describes the camera angles or perspectives employed, recorded sequentially as they appear within each segment. Common viewpoints include baseline, court-level, close-up, and bird’s-eye perspectives.
The \textit{Movements} dimension categorizes camera motion techniques utilized during each shot. These movements are documented sequentially and include static shots (no camera motion), linear motions (e.g., push-in or pull-out), and angular motions (e.g., pan, tilt, arc, or tracking shots).
Finally, the \textit{Context} dimension captures a narrative description of each shot sequence, detailing the visual storytelling intent. For example, “Close-up of scoring player (winner), followed by point replay” or “Bird’s-eye view transitioning into slow-motion ball tracking.” These narrative contexts are further classified into three overarching event types—\textit{Action}, \textit{Emotion}, and \textit{Tactic}—to facilitate structured analysis.



    


\textbf{Analysis.} 
\cref{fig:coding} (right, green portion) illustrates the distribution of camera shots across three event types: \textit{Emotion}, \textit{Tactic}, and \textit{Action}, based on our contextual annotations. \textit{Emotion} events (309 shots) dominate, highlighting broadcasters' priority in conveying the emotional atmosphere of tennis matches. These scenes target a broad audience, including casual viewers with minimal tennis knowledge, by focusing extensively on player reactions (161 instances) and capturing the overall stadium environment (99 instances). \textit{Tactic} events (175 shots) offer strategic insights aimed at more knowledgeable viewers interested in tactical analysis, primarily depicted through detailed ball-in-play movements (109 instances) and real-time statistical overlays (66 instances). \textit{Action} events (123 shots) emphasize intense gameplay sequences, showcasing dynamic ball trajectories (76 instances) and player movements (47 instances). Such sequences particularly appeal to viewers who play tennis and are invested in understanding players' physical performances and skills. Camera shot sizes and motions significantly contribute to effectively communicating these event types. Close-up shots combined with static camera movements were the most frequent (148 instances), with over half (76 instances) specifically targeting audience reactions. This choice underscores broadcasters' intent to vividly capture diverse spectator responses, thereby enhancing viewers' emotional engagement and immersing them within the match atmosphere. Medium shots combined with angular camera motions (104 instances) predominantly focus on ball-in-play events (72 instances). Such shots frequently utilize angular movements (e.g., pans and tilts) to dynamically track rapid ball trajectories and player actions, essential for clearly communicating tactical nuances and detailed gameplay developments. Wide shots, typically static or featuring minimal motion, are mainly employed to convey the stadium's overall atmosphere (26 instances), effectively establishing spatial context and emphasizing the scale of spectator participation and environmental grandeur.

Comparative analysis across the four Grand Slam tournaments further reveals distinctive stylistic variations that reflect their unique identities. The Australian Open emphasizes technical precision through frequent slow-motion replays and close-ups of footwork and stroke mechanics. The French Open integrates diverse camera angles, blending audience reactions with dynamic gameplay to heighten narrative tension. Wimbledon prefers intimate visuals characterized by slow, deliberate camera movements and frequent close-ups of player emotions, aligning with its traditional elegance. In contrast, the US Open employs rapid transitions and vibrant visuals, emphasizing energetic player actions and lively audience interactions to sustain viewer excitement and maintain narrative momentum.

\subsection{Cinematic VR Camera}\label{sec-3.2}
The audiovisual language in VR occupies a distinct intersection between traditional cinema and immersive virtual reality experiences. Unlike conventional 2D broadcasts, where viewers observe events passively on a flat screen, VR experiences situate the viewer directly within the scene, effectively turning the audience into a dynamic, moving camera. Sudden or frequent perspective shifts in this context can cause viewer disorientation and motion sickness~\cite{slater2016enhancing}. Consequently, VR camera transitions must be carefully designed, employing gradual, smooth, and predictable movements to maintain viewer comfort and immersion. To address this, we systematically analyzed existing VR narrative animations, aiming to derive guidelines suitable for VR sports broadcasting, particularly for tennis viewing.


\textbf{Data Collection.} 
To identify effective VR camera movements, we analyzed 25 cinematic VR (CVR) narrative animations from the \textit{Theater Elsewhere} application available on the Quest Store, selected for its extensive library of narrative-driven VR content. These animations, each ranging from one to ten minutes in duration, predominantly portray stories with clear moral or thematic messages, such as resilience, optimism, perseverance, and adaptability. Rather than scenic or documentary-style videos, the chosen animations prominently feature defined characters and plot-driven scenarios, closely mirroring narrative structures applicable to tennis broadcasting. Clips were segmented based on shot size and camera motion types, yielding 403 distinct camera movement instances. Although individual animations frequently reused specific camera patterns—an intentional choice to reinforce viewer immersion—our objective was to extract versatile camera movement principles suitable for adaptation into tennis-specific VR broadcasts, rather than comprehensively documenting every possible VR technique.


\textbf{Analysis.} 
Cinematic VR typically provides a seated, 180-degree viewing experience, often utilizing high frame rates to minimize discomfort. \cref{fig:coding} (left, purple portion) summarizes our analysis of shot sizes and camera motions in VR. Medium shots were most prevalent (231 instances), frequently serving as transitional views that comfortably balance narrative emphasis and viewer autonomy, allowing audiences sufficient contextual understanding while smoothly guiding their attention to key narrative details~\cite{mateer2017directing}. Close-ups (109 instances) and wide shots (63 instances) complemented medium shots, providing focused attention on crucial narrative elements and establishing broader environmental context, respectively. Regarding camera motion, linear movements dominated (199 instances), notably push-ins, pull-outs, and lateral tracking shots. These linear motions effectively and gently guided viewer attention through narrative progression without causing disorientation. Slow and deliberate linear movements enabled comfortable observation of surroundings, enhancing narrative immersion. Static shots (178 instances) facilitated concentrated observation of critical narrative points or subtle details within the scene, providing viewers with stability and reducing cognitive strain. Conversely, angular motions (26 instances), including pans and tilts, were used sparingly due to their potential to induce discomfort. These motions were carefully executed at slow speeds, restricted predominantly to close-up or medium shots, thus highlighting specific emotional or narrative focal points while minimizing viewer disorientation.

Integrating analyses from traditional 2D tennis broadcasts and cinematic VR animations, \cref{fig:coding} offers a direct comparison of camera technique usage between the two mediums. 
Each column represents the normalized frequency (with the highest value set to 100\%) of specific shot types—displayed for VR (left, purple columns) and for tennis broadcast footage (right, blue columns).
Traditional 2D broadcasts heavily utilize close-up shots to emphasize emotional engagement and detailed action through rapid cuts. In contrast, VR narratives prioritize medium shots, balancing the narrative focus with viewer autonomy, essential for maintaining immersive comfort. Moreover, traditional broadcasts frequently employ angular camera motions (e.g., rapid pans) to dynamically track fast-paced gameplay. Conversely, VR favors linear camera motions, aligning naturally with viewer movements to smoothly guide attention without overwhelming sensory processing or causing discomfort. 
By identifying and understanding these key distinctions, we mapped traditional tennis narrative elements onto VR-compatible camera movements~\cite{mills2022immersive}, laying the groundwork for our comprehensive VR tennis broadcasting design framework, detailed in the subsequent subsection.

\subsection{Design Framework}
\begin{figure}[h]
    \centering
    \includegraphics[width=\columnwidth]{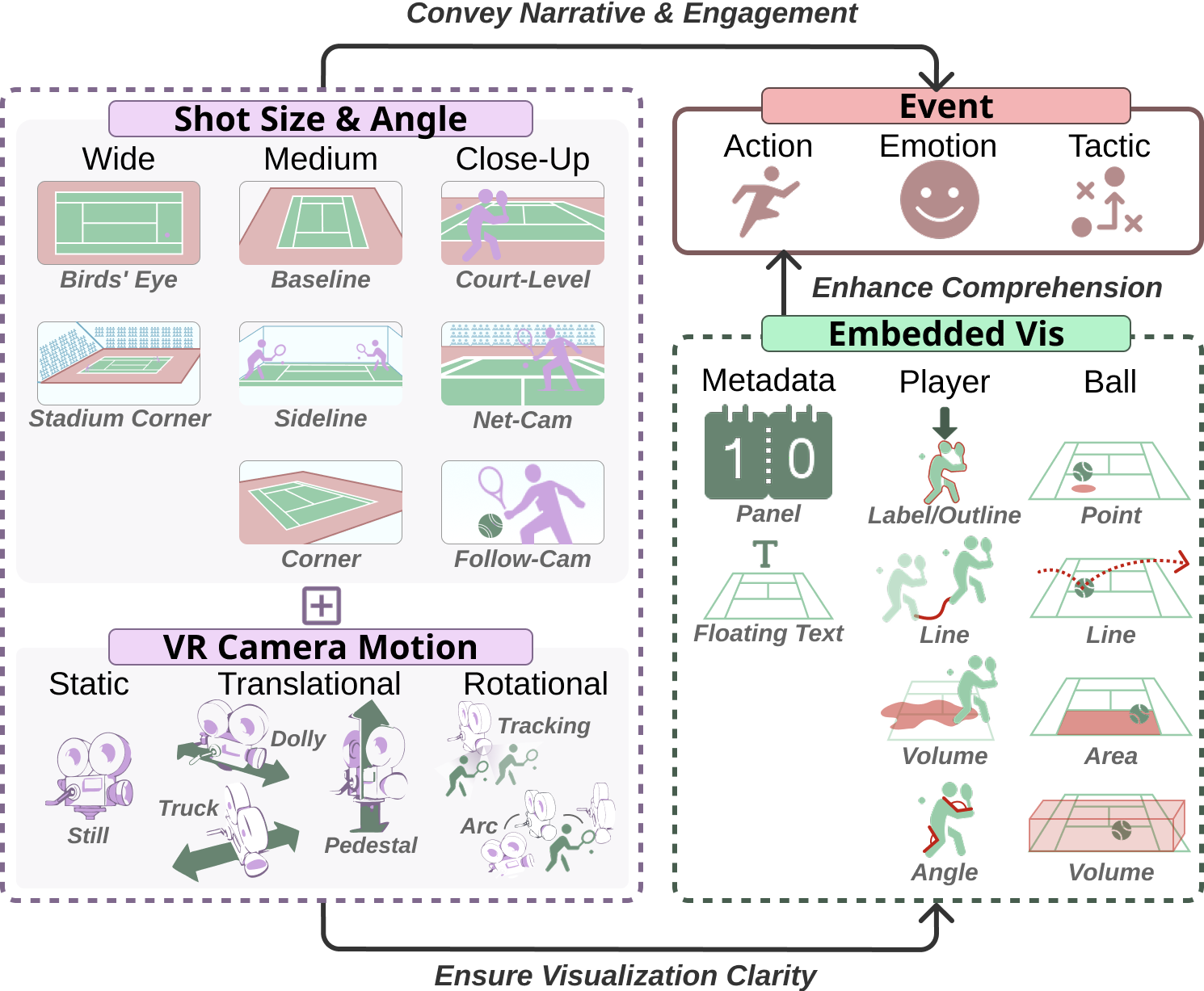}
    \caption{Our design framework integrates camera configurations—defined by shot size, angle, and motion—to convey narrative and boost viewer engagement. Embedded visualizations further complement these configurations by emphasizing key match events, thereby enriching overall comprehension and ensuring that visual data is effectively conveyed through camera techniques.}
    \label{fig:space}
    \vspace{-0.5cm}
\end{figure}

Based on insights from annotated tennis broadcasts (Section~\ref{sec-3.1}) and cinematic VR animations (Section~\ref{sec-3.2}), we propose an integrated design framework tailored specifically for VR tennis broadcasting (see \cref{fig:space}). The framework consists of four interconnected dimensions: \textit{Event}, \textit{Embedded Visualization}, \textit{Shot Size \& Angle}, and \textit{VR Camera Motion}. 
Central to the framework, the \textbf{Event} dimension guides narrative selection, directly informing the choice of camera configurations to clearly articulate narrative intent. Effective camera configurations enhance visualization clarity, and embedded visualizations subsequently reinforce viewer comprehension of event narratives. Below, we detail each dimension individually and illustrate their practical integration.


\subsubsection{Event} This dimension characterizes key gameplay scenarios that form the backbone of our narrative. We distinguished three event categories:

\textbf{Action} events capture dynamic gameplay moments, such as powerful serves, impactful rallies, and rapid exchanges. These events emphasize the physical intensity and athletic skill integral to tennis.
\textbf{Emotion} events focus on the players' reactions following critical moments, such as scoring or committing an error. Capturing celebrations, frustration, and psychological nuances, these events deepen viewer engagement by highlighting the match's emotional dimension.
\textbf{Tactic} events illustrate strategic gameplay. These events clarify tactical decision-making, strategic positioning, and nuanced match dynamics, catering to viewers interested in deeper strategic insights.
While these categories may overlap in actual gameplay, our approach distinctly separates each narrative component, facilitating focused camera techniques and visualization strategies to avoid narrative ambiguity and viewer confusion.

\subsubsection{Embedded Visualization.} This dimension complements event narratives by incorporating graphical overlays informed informed by our event annotation targets:

\textbf{Metadata} visualizations (panels, floating texts) convey essential statistical information, such as match scores and critical game highlights. Floating text visualizations further emphasize emotional significance at critical points, enhancing viewers' immediate understanding of \textit{Emotion} events.
\textbf{Player} visualizations (labels, outlines, trajectory lines, heatmaps, joint angles) explicitly illustrate annotated player actions, movements, and strategic positions. These visualizations are particularly effective in clarifying \textit{Action} and \textit{Tactic} events, enhancing comprehension of physical and strategic gameplay details.
\textbf{Ball} visualizations (trajectory lines, landing points/zones, volumetric cues) emphasize ball dynamics annotated frequently in \textit{Action} and \textit{Tactic} events. Such visual elements clearly illustrate critical tactical decisions and athletic skills, further reinforcing viewer comprehension.

\subsubsection{Shot Size \& Angle.} This dimension determines visual framing and perspective, directly informed by annotated preferences from tennis broadcasts (Section~\ref{sec-3.1}). We classify these into three categories:

\textbf{Wide Shots} provide expansive views from elevated stadium perspectives (e.g., bird's-eye, stadium corner views), frequently annotated in \textit{Tactic} events to clearly illustrate strategic player positioning, overall match context, and atmospheric emotion.
\textbf{Medium Shots} emphasize court-centric views such as the baseline (classic high-angle backcourt), sideline (emphasizing lateral dynamics), and corner views (highlighting angles and strategic court placements). Predominantly annotated for \textit{Action} events, medium shots effectively balance gameplay details and broader context, clearly depicting interactions between players and ball trajectories.
\textbf{Close-Up Shots} highlight individual players through detailed court-level perspectives (low-angle baseline), net-cam views (near-net tactical observation), and follow-cam shots (player-centric emotional detail). Annotated extensively in \textit{Emotion} events, close-up shots explicitly reveal player expressions and subtle physical movements after key points.

\subsubsection{VR Camera Motion.} This dimension defines the narrative flow and viewer perspective shifts, explicitly guided by principles identified from annotated VR narrative animations:

\textbf{Static} camera motions maintain viewer focus on annotated events where multiple simultaneous elements are critical, providing clarity without distraction from unnecessary camera movement. 
\textbf{Translational} camera motions (dolly: forward/backward, truck: lateral horizontal, pedestal: vertical) smoothly move viewers toward annotated narrative focal points. Dolly motions adjust viewing depth, highlighting close-up details or broader context. Truck motions shift attention horizontally between annotated targets. Pedestal motions vertically adjust camera height for perspective clarity, managing narrative progression and viewer spatial orientation.
\textbf{Rotational} camera motions are used to refine the viewer’s perspective. Due to VR’s unique viewing conditions—where rapid camera-centric pans, tilts, or rolls can lead to viewer disorientation—we primarily utilize object-centric rotations. These include \textit{tracking} rotations that follow key players or game elements, while \textit{arc} motions that revolve around focal points, maintaining narrative clarity and minimizing viewer disorientation.

\subsubsection{Integrated Narrative Scenarios}
To clearly illustrate how these framework dimensions integrate practically, we present scenario-based narrative examples derived from our annotated dataset (see~\cref{fig:teaser}). Each scenario corresponds explicitly to one of the three defined event categories, demonstrating how camera setups and embedded visualizations collaborate to convey the intended narrative.

For \textbf{Action events}, we highlight dynamic in-play scenarios such as powerful serves or rapid exchanges. 
Each critical point outcome, like a winning shot or a ball landing out-of-bounds, is emphasized through slow-motion replays, typically slowed to approximately 50\% speed at the moment of ball contact or bounce. 
In these scenarios, corner views are primarily chosen for these replays to simultaneously display the player's precise stroke mechanics and the ball’s trajectory clearly. 
Additionally, for specific dynamic situations such as net approaches, a \textit{Net-Cam} perspective is employed to closely capture the moment of impact. 
In these \textit{Action} sequences, camera movements remain predominantly \textit{static} or minimally adjusted to ensure clear viewing of fast-moving gameplay elements, minimizing viewer disorientation and enhancing narrative clarity.
In \textbf{Tactic events}, we emphasize strategic gameplay elements like shot selection, player positioning, and rally patterns. 
\textit{Wide} or \textit{Medium} shots, such as bird’s-eye or baseline views, provide comprehensive spatial context. 
Static visualizations, including color-coded heatmaps and statistical overlays derived from annotated positional data, can clearly reveal the underlying strategic patterns. 
Camera movements such as subtle \textit{dolly}, \textit{pedestal}, and \textit{arc} motions adaptively guide viewer attention toward specific annotated strategic points. 
For long rallies, baseline views are used to maintain consistent perspective, whereas sideline or judge views reduce excessive viewer head movement. 
Short tactical plays utilize dedicated corner angles to explicitly highlight serve placement and critical tactical details.
For Emotion events, which highlight annotated emotional responses from players following key match points, we primarily employ \textit{Close-Up Follow-Cam} shots paired with smooth tracking camera movements. 
These shots explicitly capture detailed player expressions and nuanced physical reactions, directly conveying emotional intensity. 
Additionally, broader \textit{Wide Shots} combined with subtle dolly or pedestal motions effectively emphasize the overall stadium atmosphere, visually contextualizing crowd reactions and environmental excitement. 
Dynamic embedded visualizations, such as floating text annotations ("game point," "set point") and celebratory icons, further enhance emotional highlights and amplify their narrative significance.
Collectively, these scenario integrations effectively demonstrate how our design framework translates annotated insights into immersive and coherent VR tennis narratives.

%% file: Section/4-Method.tex
\begin{figure*}[htb]
     \includegraphics[width=\linewidth]{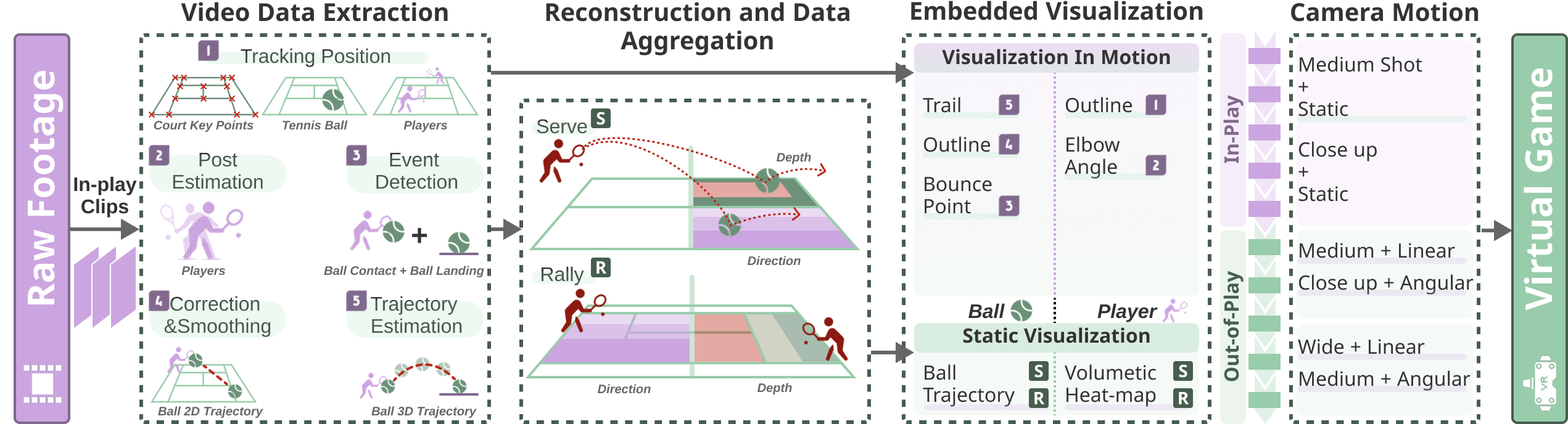}
    \caption{Overview of our system pipeline for converting raw broadcast footage into an immersive VR tennis experience. Dynamic tracking data, extracted from the footage, feeds into the visualization in Motion module to overlay real-time cues—such as ball trajectories and impact highlights. Simultaneously, simulated game data is aggregated to compute event metrics (e.g., ball bounce area placements), which are presented as static visualizations. Finally, adaptive camera motions integrate these components, guiding viewer attention and constructing a coherent narrative.
    }
     \label{fig:pipeline}
     \vspace{-0.5cm}
\end{figure*}

\section{System Design and Implementation}\label{sec-4}
Building on our design exploration in Section \ref{sec-3}, we propose and implement \emph{Beyond the Broadcast}, a VR tennis viewing system that seamlessly integrates tailored camera techniques with embedded visualizations. Our system comprises four primary modules: \textit{Video Data Extraction}, \textit{Reconstruction Scene}, \textit{Embedded Visualization}, and \textit{Camera Motion Planning} (see~\cref{fig:pipeline}). In the following subsections, we provide a detailed description of each module.

\subsection{Video Data Extraction}
To accurately reconstruct 3D tennis match from monocular broadcast videos (25 fps) sourced from platforms like \textit{YouTube}~\cite{yuan2023learning, zhang2021vid2player}, we developed a semi-automatic pipeline leveraging state-of-the-art computer vision techniques. Initially, the raw footage is segmented into discrete in-play clips~\cite{de2023automatic}. Subsequent automated modules detect court keypoints, track player poses, capture ball trajectories, and identify critical match events such as ball bounces and player-ball contacts. Manual annotation is also employed to refine ball height estimations, manage rare occurrences (e.g., net-cord incidents), and correct data outliers.

\textbf{Tracking Position and Pose Estimation.}
Accurate court line detection is fundamental to mapping player and ball positions onto a consistent reference court. We employ a pre-trained neural network to detect 14 court keypoints per frame. These detected points are combined with known court reference coordinates to compute a homography matrix, enabling precise alignment between screen and court space. Our enhanced court detection module achieves a precision of 96.3\%, an accuracy of 96.1\%, and a median positional error of only 1.83 pixels.
To obtain the ball position in each frame, we leverage the TrackNet deep learning model~\cite{huang2019tracknet}, which processes three consecutive frames to infer the ball trajectory. This approach is particularly effective in addressing the challenges posed by small, blurry, or occluded ball images in broadcast videos. Our implementation yields an F1-score of 89.9\%, surpassing the original TrackNet benchmark (F1-score 84.3\%).
Player detection employs a custom-trained YOLOv8 model~\cite{reis2023real}, followed by pose estimation using GVHMR~\cite{shen2024world}, which calculates 23 joint parameters capturing rotational and positional data. This pose estimation represents the most computationally intensive step, requiring approximately 100 ms to process a 503-frame segment (\textasciitilde20 seconds) using an NVIDIA RTX 4090 GPU. Pose outputs are exported in FBX format for subsequent 3D reconstruction.

\textbf{Event Detection.}
Accurate event detection is essential for segmenting match footage and enabling detailed trajectory reconstruction. Ball bounce events, indicating the precise moment of ground contact, are detected using a CatBoost classifier~\cite{prokhorenkova2018catboost}, analyzing ball positions over a sequence of 20 frames. Evaluation metrics demonstrate a precision of 74.4\% and recall of 73.2\% for exact frame matching. Allowing a ±1 frame tolerance improves precision to 91.1\%. Additionally, ball-player contact events are identified using a model proposed by Hong et al.~\cite{hong2022spotting}, achieving similar robust performance within a ±1 frame tolerance. These steps ensure reliable event detection, which is critical for accurate trajectory estimation.

\textbf{Correction and Smoothing.}
After extracting coordinates, we obtain two-dimensional time series data representing the positions of players and the ball for each clip.
Minor perturbations and occasional missing player data (due to occlusions) are corrected using k-nearest neighbors interpolation (k=5), followed by smoothing with a 5-frame moving average filter. Since coordinates are projected from 2D broadcast footage into 3D space, resolution limitations can introduce jitter. A resolution-based stabilization step further smooths player trajectories, significantly reducing artifacts.
For ball position data, rapid motion and altitude-induced projection errors cause significant jitter, particularly along the X-axis. To mitigate this, a two-stage physical validation approach is implemented. First, ball-ground contact frames (\(Z=0\)) establish a reliable Y-axis baseline reference, allowing outlier removal due to occlusions. Second, at ball-player impact frames, the ball's Y-coordinate is substituted with the player's foot-contact position to minimize projection inaccuracies.
We subsequently reconstruct ball trajectories using segmented kinematic modeling between keyframes, assuming constant-velocity linear motion (ignoring air resistance). The velocity components \((v_x, v_y)\) are calculated as:
\[
(v_x, v_y) = \left(\frac{\Delta s_x}{\Delta t}, \frac{\Delta s_y}{\Delta t}\right),
\]
where \(\Delta s_x\) and \(\Delta s_y\) represent positional changes between consecutive keyframes, and \(\Delta t\) is the elapsed time. This approach effectively reconstructs accurate two-dimensional trajectories, minimizing altitude-induced projection errors. After automated detection, correction, and smoothing, extracting and refining ball and player positions for a 503-frame (\textasciitilde20 seconds at 25 fps) video segment takes approximately 45 ms on an NVIDIA RTX 4090 GPU, resulting in a total of about 145 ms when combined with the pose estimation module.

\textbf{Trajectory Estimation.}
Due to the inherent limitations of monocular video, direct 3D ball trajectory estimation is not feasible. Therefore, we adopt a kinematic reconstruction approach guided by manually annotated keyframes. Ball initial heights and rotation types (topspin or backspin) at contact points are explicitly labeled. Assuming uniformly accelerated linear motion along the vertical axis (\(Z\)), we use accelerations of \(9.81\,\mathrm{m/s^2}\) for topspin and \(10.81\,\mathrm{m/s^2}\) for backspin shots. For specific events, such as net-cord incidents, we explicitly mark keyframes with predetermined heights (e.g., \(Z=0.9\,m\) at net impact, \(Z=0\,m\) at ground contact).
Segmented kinematic modeling between annotated keyframes reconstructs vertical trajectories using the following kinematic relation:
\[
h_i = h_{i-1} + v_i t_i + \frac{1}{2} a_i t_i^2, \quad \text{for } i=1,\dots,n,
\]
where \(h_i\) and \(h_{i-1}\) denote ball heights at consecutive keyframes, \(a_i\) is the acceleration during the interval, and \(t_i\) is the elapsed time between frames. This formulation ensures accurate estimation of intermediate velocities \(v_i\), enabling reliable reconstruction of the ball’s complete 3D trajectory from monocular footage.

\subsection{Reconstruction Scene}
Using the positional data extracted from the previous video processing stage, we reconstruct a detailed 3D representation of the tennis match. This reconstructed scene precisely captures the movements and positions of players and the ball, serving as a foundation for comprehensive tactical analyses. The tennis court is divided into directional and depth-based zones, separately identified for serves and rallies. Ball bounces and player-ball contact events are systematically logged within these zones, enabling precise event metrics. 
These metrics—expressed either as percentages or counts—can be calculated from the match's start or from the beginning of the current game, facilitating both real-time insights and detailed retrospective analysis. To evaluate the runtime performance of our reconstruction pipeline, we conducted tests on a laptop equipped with an NVIDIA RTX 4060 GPU. The average runtime achieved is approximately 120 frames per second (FPS), demonstrating the system's suitability for real-time VR tennis broadcasting scenarios.

\subsection{Embedded Visualization}
Our embedded visualization module integrates graphical overlays directly into the VR broadcast, enhancing both tactical insight and game understanding. Inspired by advanced visualization techniques utilized in modern sports broadcasts—such as floating text and court overlays featured at the \textit{2025 Australian Open}—as well as popular tennis simulation games like \textit{Top Spin 2K25} and \textit{AO Tennis 2}, our system combines dynamic and static visualization approaches. Specifically, dynamic visualizations present animated overlays of ball trajectories, player movements, and real-time match statistics, while static visualizations deliver aggregated tactical insights, such as volumetric heatmaps and comprehensive shot statistics, reinforcing overall narrative coherence.

\textbf{Dynamic visualizations} offer real-time overlays that facilitate intuitive interpretation of in-play actions and detailed slow-motion replays~\cite{zhao2012matching, chen2015novel}. These visual cues include trajectory trails, highlight outlines that mark critical ball bounces and player impacts, and specific annotations such as elbow angles during key strokes. Additional aids include directional serve indicators, floating text labels (e.g., "game point," "set point"), and real-time shot counts, all enhancing the immediate understanding of match progression. 
\textbf{Static visualizations} leverage data from reconstructed 3D scenes to present macro-level overviews of match dynamics. Examples include full-court, color-coded ball trajectory maps and player positional volumetric heatmaps, which elucidate underlying tactical patterns and strategic insights. These visualizations complement dynamic overlays by providing a holistic view of gameplay.
Together, these visualization methods enable our system to effectively communicate complex tactical and emotional narratives within the VR tennis broadcast environment.

\subsection{Camera Motion Planning}
Our camera motion planning module orchestrates viewpoint transitions to deliver a coherent and engaging narrative. Drawing insights from the event segmentation detailed in Section~\ref{sec-3}, the module strategically selects camera configurations tailored to each match scenario. This ensures visual clarity, narrative coherence, and viewer comfort.

For \textbf{In-Play segments}, the module primarily employs medium shot perspectives (e.g., baseline views) combined with static camera positions. This approach maintains viewer focus on immediate gameplay dynamics, reducing unnecessary visual distractions. Conversely, during \textbf{Out-of-Play segments}, the module leverages carefully designed camera motions to provide analytical insights and detailed replays. Medium shots combined with smooth linear (dolly) camera movements offer gradual zoom-ins for detailed analysis, while close-up shots integrated with subtle rotational (arc) movements highlight critical tactical elements. Wide shots, paired with linear motions, provide comprehensive overviews of the court, and medium shots with gentle angular adjustments emphasize specific tactical contexts. This adaptive camera motion strategy minimizes viewer disorientation, addressing common challenges in VR viewing experiences, and reinforces overall narrative clarity. To precisely integrate these tailored camera motions with embedded visualizations during both In-Play and Out-of-Play scenarios, we employ Unity’s Timeline system for accurate and flexible arrangement.

%% file: Section/5-Evaluation.tex
\section{User Study}\label{sec-5}
This section describes our controlled user study designed to evaluate the overall effectiveness of \emph{Beyond the Broadcast} in terms of comprehension, engagement, immersion, and viewing comfort. We present details on participant recruitment, study design, materials, procedures, and evaluation measures. 

\subsection{Participants}
We recruited 16 participants (8 male, 8 female; P1–P16), aged between 20 and 34 years ($\mu = 25.88$, $\sigma = 3.52$). All participants were graduate students majoring in data visualization, applied statistics, or software engineering. Regarding their familiarity with tennis, three participants reported minimal prior exposure; eleven described themselves as casual fans who occasionally watch major tournaments (e.g., Grand Slams or the Olympics) but have never attended live matches; and two participants identified as avid fans with experience attending live events. Participants' experience with VR varied: four had no prior VR exposure, eight had limited experience, and four were experienced VR users. While some participants reported mild discomfort related to 3D environments—such as occasional vertigo during first-person shooter games—these symptoms were minor and did not significantly influence their participation or overall study experience. One participant noted a mild fear of heights but reported no considerable distress during the study. All participants had normal or corrected-to-normal vision, no prior involvement in the system design, and received \$7 USD as compensation for their participation.

\subsection{Study Design}
We established three experimental conditions within an immersive VR environment, each featuring a distinct set of tennis scenarios. Specifically, we extracted 18 individual points from the 2019 Wimbledon Men's Final and grouped them evenly into three clips, each containing six points. These points were carefully selected using event segmentation principles, ensuring a balanced representation of key tennis moments such as winning shots, errors, extended rallies, and notable tactical plays (e.g., one-two punch). Each clip's total duration ranged from one to two minutes, preserving the dynamic and authentic nature of live tennis broadcasts. The three experimental conditions are defined as follows: In the \textbf{Baseline} condition, participants viewed simulated original broadcast settings without embedded visualizations or dynamic camera movements. A scoreboard placed in the corner mimicked the passive and fixed-viewing perspective typical of live tennis arenas. The \textbf{Visualization Only} condition introduced dynamic embedded visualizations, emphasizing tactical insights, player movements, and ball trajectories while keeping the camera viewpoint static. Finally, the \textbf{Full Integration} condition combined both dynamic and static embedded visualizations with adaptive camera motions guided by CVR shot techniques, crafting a coherent narrative experience.

To minimize potential order effects, the sequence of conditions presented to each participant was randomized. Recognizing the varied levels of VR experience among participants, we carefully designed the VR content to reduce potential motion sickness and discomfort. Participants were seated during the study, and primary visual content was confined to a frontal viewing range of 180 degrees. This approach controlled biases arising from unfamiliarity with VR hardware, enabling us to effectively focus on assessing viewer responses specifically to the visualization design and camera motion strategies. Furthermore, the VR experience was augmented with original match audio—including crowd ambiance, commentary, and player interactions—to authentically replicate the atmosphere of a live tennis event, thus enhancing ecological validity.

\begin{figure}[h]
    \centering
    \includegraphics[width=\columnwidth]{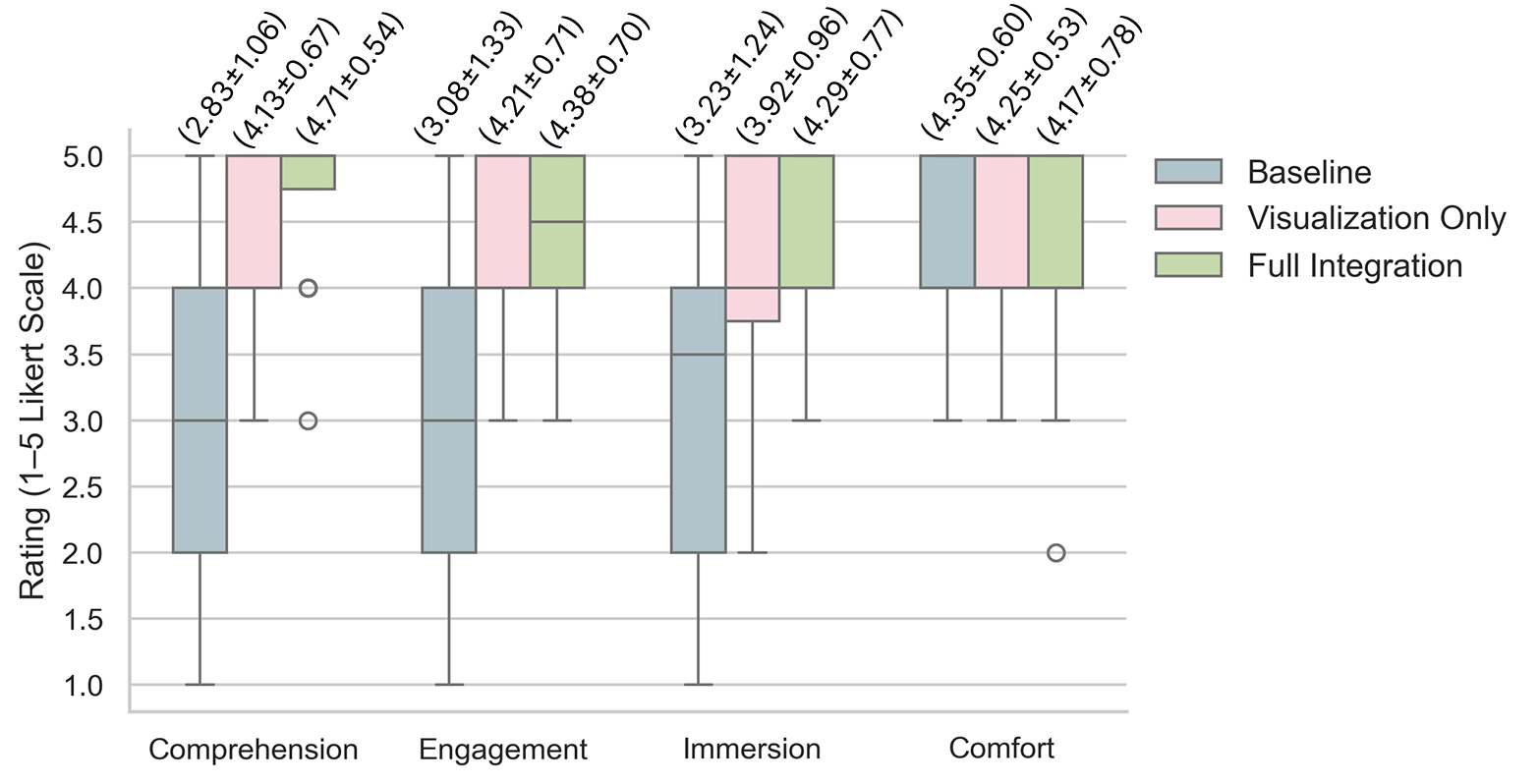}
    \caption{Participants' ratings (1–5 Likert scale) for Comprehension, Engagement, Immersion, and Comfort across three conditions: Baseline, Visualization Only, and Full Integration. Results indicate notable improvements in Comprehension, Engagement, and Immersion with embedded visualizations and adaptive camera movements, while Comfort remains consistently high across conditions.}
    \label{fig:us}
    \vspace{-0.5cm}
\end{figure}

\subsection{Materials}
\begin{figure*}[htb]
     \includegraphics[width=\linewidth]{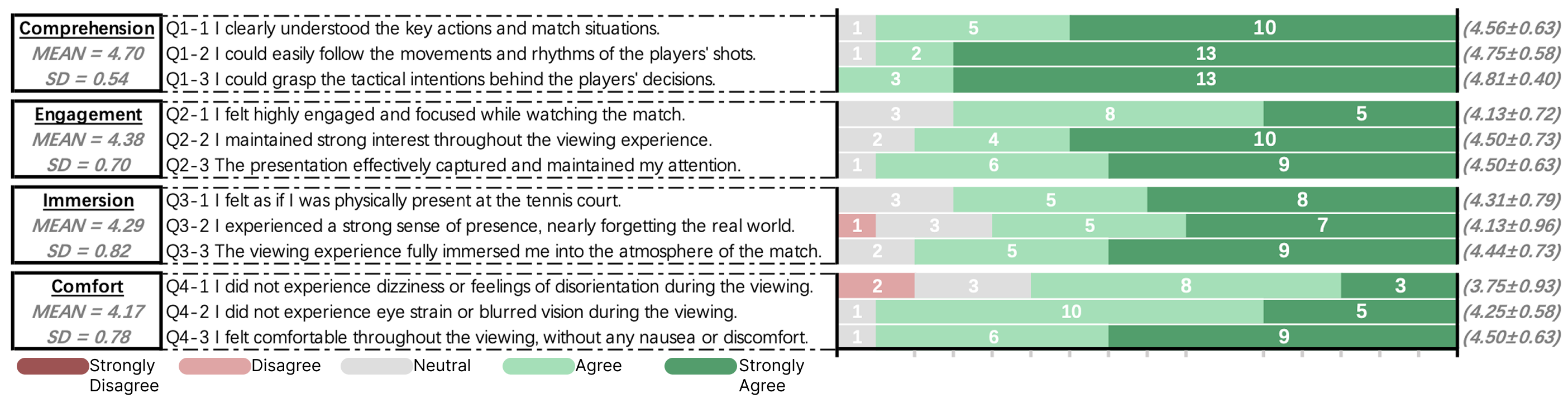}
    \caption{Detailed survey responses for the Full Integration condition, showing participants' agreement levels (1–5 Likert scale) with individual statements related to Comprehension, Engagement, Immersion, and Comfort. Mean ($\mu$) and SD ($\sigma$) for each item are shown on the right.}
     \label{fig:us2}
     \vspace{-0.5cm}
\end{figure*}

The study was conducted using an Oculus Quest 3 Head-Mounted Display (HMD) connected to a high-performance laptop equipped with an RTX 4060 GPU. We assessed participants' experiences across four key dimensions—\textit{Comprehension}, \textit{Engagement}, \textit{Immersion}, and \textit{Comfort}—using a customized questionnaire adapted from the Film Immersive Experience Questionnaire (FilmIEQ)~\cite{rigby2019development} and the Virtual Reality Sickness Questionnaire (VRSQ)~\cite{kim2018virtual}. Each dimension comprised three targeted questionnaire items, as detailed in~\cref{fig:us2}. Specifically, the \textit{Comprehension} assessed narrative understanding, focusing on participants' clarity in identifying key match events \textit{(Q1-1)}, tracking players' movements and shot rhythms \textit{(Q1-2)}, and interpreting tactical strategies within the match context \textit{(Q1-3)}. Given varying levels of tennis familiarity among participants, these items emphasized general scenario comprehension and rhythm-following rather than advanced tactical knowledge. The \textit{Engagement} dimension evaluated narrative captivation and cognitive involvement~\cite{pillai2019grammar}, examining overall attentiveness \textit{(Q2-1)}, sustained interest \textit{(Q2-2)}, and efficacy of the presentation in capturing attention \textit{(Q2-3)}. The \textit{Immersion} measured perceptual immersion~\cite{skarbez2017survey}, addressing participants' subjective sense of spatial presence on the virtual court \textit{(Q3-1)}, their detachment from real-world surroundings \textit{(Q3-2)}, and depth of immersion into the virtual tennis atmosphere \textit{(Q3-3)}. Lastly, the \textit{Comfort} evaluated potential VR-induced physical discomfort~\cite{tian2022review} by assessing the absence of dizziness or disorientation \textit{(Q4-1)}, oculomotor issues such as eyestrain or blurred vision \textit{(Q4-2)}, and overall physical comfort without excessive fatigue or nausea \textit{(Q4-3)}. To facilitate clear comparisons across conditions, the questionnaire was intentionally streamlined rather than employing the complete FilmIEQ and VRSQ scales. Qualitative interviews supplemented these measures to offer deeper insights into participants' experiences and perceptions.



\subsection{Procedure}
The study began with an introduction to the experimental protocol, followed by informed consent and a demographic survey. Participants were asked to assume the role of tennis viewers in VR. Initially, participants spent approximately one minute familiarizing themselves with the HMD in a virtual tennis court. Subsequently, participants viewed three VR video clips presented in randomized order without interruption. Immediately after each clip, participants verbally recalled key events and tactical details, providing a qualitative measure of comprehension. Following this, they completed a Likert-scale questionnaire. Participants were allowed to revisit and adjust their ratings from prior clips after experiencing subsequent conditions. Five-minute breaks were provided between clips to reduce fatigue and monitor for VR-induced discomfort. Finally, a 15-minute semi-structured interview was conducted to further explore participants' perceptions of the visualizations and camera motions. Each session lasted approximately 40 minutes, with at least six cumulative minutes of VR exposure.


\subsection{Results and Findings}
Overall participant ratings across conditions and evaluation dimensions are summarized in~\cref{fig:us}, while~\cref{fig:us2} details participant responses to individual questionnaire items under the \textit{Full Integration} condition.

\textbf{Enhanced attention management and improved comprehension.}
Participants reported significantly improved comprehension across experimental conditions \textit{(Baseline:$\mu=2.8$, Visualization Only:$\mu=4.1$, Full Integration:$\mu=4.7$)}. In the \textit{Baseline} condition, participants faced substantial challenges tracking rapid ball movements and simultaneous player actions. As P4 commented, \textit{``I struggled even to consistently follow the ball due to multiple simultaneous actions that scattered my attention.''} Dynamic visualizations noticeably reduced cognitive load by clearly highlighting key events, enabling better allocation of attention between ball trajectories and player actions. As P8 remarked, \textit{``The visual cues significantly eased my cognitive load, allowing my attention to shift comfortably from solely tracking the ball to focusing on player strategies.''} Further, adaptive camera movements in the \textit{Full Integration} condition strategically guided viewer attention toward critical match events, minimizing the effort required from viewers to identify key moments independently. P11 explained, \textit{``Adaptive camera framing clearly emphasized crucial moments, helping me immediately grasp the significance without actively shifting attention.''} Nevertheless, certain participants identified instances of visual clutter during extended rallies, which somewhat overwhelmed their attention. For example, P15 noted, \textit{``In long rallies, static trajectory overlays sometimes became visually overwhelming without additional context.''} Overall, these findings highlight the benefit of progressively integrating visual information, starting minimally and gradually increasing detail to align with viewer attention management capabilities.

\textbf{Strengthened narrative engagement with perceptual immersion considerations.}
Participants reported increased narrative engagement \textit{(Baseline:$\mu=3.1$, Visualization Only:$\mu=4.2$, Full Integration:$\mu=4.4$)} alongside enhanced perceptual immersion \textit{(Baseline:$\mu=3.2$, Visualization Only:$\mu=3.9$, Full Integration:$\mu=4.3$)}. Dynamic visualizations and adaptive camera techniques effectively heightened narrative captivation by clearly framing and guiding viewers toward pivotal match events. P6 highlighted, \textit{``The shifting camera angles made the match feel like an engaging story unfolding in front of me.''} Despite these narrative benefits, perceptual immersion revealed certain trade-offs: participants occasionally found embedded visualizations overly conspicuous, slightly diminishing realism. As P10 observed, \textit{``Some visualizations stood out too prominently, momentarily making the scene feel artificial and disrupting my immersion.''} Additionally, participants indicated lower immersion in the \textit{Visualization Only} condition due to static camera positioning. P5 stated, \textit{``Visualizations alone felt informative but less immersive without dynamic camera movements, reducing the sensation of truly being present at the match.''} Conversely, several participants appreciated the \textit{Baseline} condition's simplicity, viewing it as more authentic to an actual tennis environment, though lacking crowd atmosphere. P8 mentioned, \textit{``Baseline felt closest to a live match experience, but sometimes it felt isolated without crowd noise or atmosphere.''} Participants suggested enhancing social presence through elements like spatial audio, realistic crowd reactions, commentary, which could further elevate overall perceptual immersion. Overall, these insights emphasize the critical balance between enhancing narrative engagement and maintaining perceptual realism~\cite{jung2021perspective} and social presence for optimal immersive experiences.

\textbf{Balanced viewing comfort with considerations for presence and cybersickness.}
Participants maintained high comfort ratings across conditions \textit{(Baseline:$\mu=4.4$, Visualization Only:$\mu=4.3$, Full Integration:$\mu=4.2$)}. Slow linear camera motions and gentle rotations around objects were highly praised, facilitating comfortable viewing without notable discomfort. Participant P3 mentioned, \textit{``The slow forward camera movement allowed me to comfortably focus on key details, and gentle rotations provided clear perspectives without discomfort.''} However, certain camera movements, especially vertical transitions and abrupt viewpoint cuts to overhead perspectives, occasionally caused mild disorientation or dizziness. For instance, P7 noted, \textit{``The downward camera movement created a brief sense of dizziness and disorientation.''} Additionally, quick cuts to aerial views disrupted the spatial orientation. 
Increased frequency of camera motion raised cybersickness concerns, with participants indicating in interviews that one to two movements per point were comfortably acceptable. Sudden shifts in viewpoint were also reported to disrupt attention, pulling viewers away from key elements. This aligns with existing research~\cite{weech2019presence} highlighting a negative relationship between presence and cybersickness. Thus, moderating camera motion frequency and avoiding abrupt perspective changes are essential to optimize viewer comfort.

Overall, the findings from the user study confirm that integrating embedded visualizations with adaptive camera techniques substantially improves viewer comprehension, narrative engagement, and perceptual immersion, without significantly compromising comfort. These results validate the effectiveness of our design framework for immersive VR sports broadcasting, demonstrating clear potential for application and further refinement.

%% file: Section/7-Limitation.tex
\section{Discussion}\label{sec-7}
In this section, we revisit our design framework and VR broadcasting system, highlighting their general applicability, discussing critical design trade-offs, and identifying opportunities to enhance viewer experience through future research.

\textbf{Generalizability to other sports domains.}
Our proposed framework provides foundational guidelines that can be readily adapted to other racket sports (e.g., badminton, table tennis), which similarly employ stable, full-court broadcasting perspectives as primary input. These sports share comparable narrative structures, player dynamics, and ball trajectories, enabling efficient translation of our design framework and data-driven pipeline with modest adjustments for court dimensions and sport-specific dynamics. However, extending our approach to complex team sports such as basketball or soccer introduces substantial challenges, including multi-player interactions, frequent occlusions, and intricate tactical considerations. Addressing these challenges will require advanced multi-agent tracking techniques, occlusion-aware reconstruction approaches, and sport-specific camera grammars tailored explicitly to the dynamics and strategic complexities inherent in team competitions. Future research can explore these enhancements, ultimately broadening the applicability and narrative potential of our VR broadcasting methodology across diverse sporting contexts.

\textbf{Trade-offs in visualization complexity and viewer attention.}
In sports broadcasting, visualizations serve primarily as supplementary aids to enhance viewer understanding without overshadowing the core dynamics of the game itself.
Our approach employs dynamic visualizations for highlighting critical in-game moments, while static visualizations support more detailed post-game analyses.
Although dynamic visualizations effectively guide viewer attention by positioning cues close to key action areas and limiting their duration, static visualizations—such as cumulative ball trajectories—can inadvertently introduce visual clutter. 
This clutter may overwhelm viewers, reducing narrative clarity and cognitive effectiveness. 
Addressing this trade-off requires refining visualization techniques—such as providing concise summaries, higher-level aggregations, or incorporating expert commentary—to bridge the gap between detailed analytical content and audience-friendly presentation.
Future research should further explore these design considerations to effectively balance visualization complexity with viewer engagement and cognitive clarity.

\textbf{Balancing viewer agency and guided narrative.}
Our current study employs a passive viewing approach, which tends to result in lower presence and increased motion sickness compared to interactive viewing experiences. Interactive designs, such as those used in e-sports broadcasts with real-time data overlays and multiple camera perspectives, have shown potential to enhance immersion and reduce viewer discomfort~\cite{kokkinakis2020dax,sell2015sports}. However, previous studies caution that excessively complex controls can negatively impact viewer engagement and satisfaction~\cite{aksun2022enhancing,carlsson2015designing}. Our participants expressed diverse preferences regarding user agency: some favored minimal input, preferring automatic visual cues at key moments, while others desired greater control over displayed content. This highlights a need for flexible interaction models that effectively balance automation and user agency. Future research could explore lightweight interaction modalities (e.g., gaze-based, gesture recognition, or voice commands) that allow context-aware control without disrupting the dynamic nature of sports viewing. We envision incorporating intelligent production agents to provide personalized viewing experiences, adaptively managing visual and camera cues to optimize viewer engagement and comfort.


{\textbf{Limitations and future work.}
Although we received encouraging feedback, our user study was limited by a relatively small and homogeneous sample. Future studies should involve a more diverse participant base in age, tennis expertise, and VR familiarity to better understand demographic influences on immersion and comprehension, thus enabling iterative refinement of our system. Additionally, minor camera jitter caused by slight head movements during HMD usage was noted; implementing advanced stabilization techniques could enhance viewer comfort and visual clarity. Furthermore, the accuracy of player motion reconstruction and emotional expressiveness is constrained by broadcast video limitations. Future work can employ physics-based methods~\cite{yuan2023learning, ertner2024synthnet} and richer training datasets to improve visual realism. Additionally, our current system involves considerable processing time and manual annotations, restricting real-time broadcasting capabilities—a limitation not yet resolved by current volumetric reconstruction methods. Finally, multimodal elements (spatial audio, dynamic commentary, subtle haptics) could further enhance viewer immersion and replicate the social atmosphere of live tennis events.

%% file: Section/8-Conclusion.tex
\section{Conclusion}\label{sec-8}
This paper presents \emph{Beyond the Broadcast}, an VR tennis broadcasting system integrating embedded visualizations with adaptive cinematic VR camera techniques. Our approach effectively merges traditional broadcast principles with immersive VR design methodologies. A user study demonstrates enhancements in viewer engagement, comprehension and perceptual immersion without compromising comfort. Future research will focus on refining visualization clarity, optimizing adaptive camera controls, exploring personalized viewer interactions, and extending this immersive broadcasting framework to other sports.